\documentclass[graybox, natbib, nosecnum, twocolumn]{svmult}
\usepackage{ifthen} \newboolean{isSpringer} \setboolean{isSpringer}{true}

\usepackage[utf8]{inputenc}

\usepackage{mathptmx}       
\usepackage{helvet}         
\usepackage{courier}        
\usepackage{type1cm}        

\usepackage{makeidx}         
\usepackage{graphicx}        
\usepackage{multicol}        
\usepackage[bottom]{footmisc}
\usepackage[normalem]{ulem}	
\usepackage{soul}   

\usepackage[utf8]{inputenc}
\usepackage{amsfonts,amsmath,amssymb} 
\usepackage{eurosym} 
\usepackage[defblank]{paralist} 
\usepackage[obeyspaces]{url} 
\usepackage[verbatim]{svn-multi}
\usepackage[dvipsnames]{xcolor}
\usepackage{xspace}
\usepackage{seclabel}
    \seclabel{sec:definition}{Definition}
    \seclabel{sec:background}{Background}
    \seclabel{sec:consistency-models}{Consistency Models}
    \seclabel{sec:basic-concepts}{Basic Concepts}
    \seclabel{sec:definitions}{Definitions}
    \seclabel{sec:results}{Fundamental results}
    \seclabel{sec:trade-offs}{Trade-offs}
    \seclabel{sec:common-models}{Common models}
    \seclabel{sec:dimensions}{A three-dimensional view of data consistency}

\usepackage{crossref} 
\svnidlong%
	{$HeadURL: svn+ssh://shapiro@scm.gforge.inria.fr/svn/regal/trunk/papers/2017/shapiro/encyclopedia-consistency/main.tex $}
	{$LastChangedDate: 2018-03-19 18:27:45 +0100 (Lun, 19 mar 2018) $}
	{$LastChangedRevision: 7314 $}
	{$LastChangedBy: shapiro $}

\usepackage{graphicx} 
\usepackage{color}  
\usepackage{tikz}
\usetikzlibrary{decorations.pathreplacing,positioning,automata,calc}
\usepgflibrary{shapes.symbols} 
\usetikzlibrary{shapes.symbols} 

\newcommand{\inc}[1]{\ensuremath{\mathit{#1\texttt{++}}}}
\newcommand{\dec}[1]{\ensuremath{\mathit{#1\texttt{--}}}}
\newcommand{\aput}{\ensuremath{\mathit{put}}}
\newcommand{\get}{\ensuremath{\mathit{get}}}
\newcommand{\nil}{\ensuremath{\mathit{nil}}}

\newcommand{\readFrom}{\ensuremath{\xrightarrow{\text{wr}}}}

\ifthenelse{\boolean{isSpringer}}{
}{
   \usepackage{leading} \leading{6mm}
   \usepackage[authoryear,round]{natbib}
   \setcounter{secnumdepth}{0}
   \usepackage{fancyhdr}
   \pagestyle{fancy} \fancyhf{} 
   \cfoot{\thepage{}}
   \bibpunct{(}{)}{;}{a}{}{,} 
}

\let\cite=\citep
\let\citetable=\citet

\usepackage[
  draft,
  colorlinks = true,
  linkcolor = black,
  urlcolor  = black,
  citecolor = black,
  anchorcolor = black]{hyperref}

\begin{document}

  \author{
    Marc Shapiro \\
    {\small Sorbonne-Université \& Inria, Paris, France}
    \and
    Pierre Sutra\\
    {\small Télécom SudParis, Évry, France}
  }

  \ifthenelse{\boolean{isSpringer}}{
  \title*{%
    Database consistency models%
  }
  \institute{Marc Shapiro
    \at Sorbonne-Universités-UPMC-LIP6 \& Inria Paris,
    \email{http://lip6.fr/Marc.Shapiro/}}
  \institute{Pierre Sutra
    \at Télécom SudParis,
    \email{pierre.sutra@telecom-sudparis.eu}}%
  \authorrunning{Shapiro \& Sutra}
}{
  \title{%
    Database consistency models%
  }
  \date{18 March 2018}
}

\maketitle
\section{Synonyms}

Consistency model, data consistency, consistency criterion, isolation
level.

The distributed systems and database communities use the same word, {consistency}, with different meanings.
Within this entry, and following the usage of the distributed algorithms
community, ``{consistency}'' refers to the observable behaviour of a
data store.

In the database community, roughly the same concept is called
``isolation,'' whereas the term ``consistency'' refers to the property that
application code is sequentially safe (the C in ACID).

\section{Definition}

A data store allows application processes to put and get data from a
shared memory.
In general, a data store cannot be modelled as a strictly sequential process.
Applications observe non-sequential behaviours, called anomalies.
The set of possible behaviours, and conversely of possible anomalies,
constitutes the \emph{consistency model} of the data store.

\section{Overview}

\subsection{Background}

A \emph{data store}, or database system, is a persistent shared memory
space, where different client application processes can store data items.
To ensure scalability and dependability, a modern data store distributes
and replicates its data across clusters of \emph{servers} running in
parallel.
This approach supports high throughput by spreading the load, low
latency by parallelising requests, and fault-tolerance by replicating
data and processing; system capacity increases by simply adding more
servers (scale-out).

Ideally, from the application perspective, data replication and distribution
should be transparent.
Read and update operations on data items would appear to execute as in a
single sequential thread; reading a data item would return exactly the
last value written to it in real time; and each application transaction
(a grouping of operations) would take effect atomically (all at once).
This ideal behaviour is called \emph{strict serialisability}, noted SSER
\cite{Papadimitriou:1979}.

In practice, exposing some of the internal parallelism to clients
enables better performance.
More fundamentally, SSER requires assumptions that are unrealistic at
large scale, such as absence of network partitions (see
\secref{sec:results} hereafter).
Therefore, data store design faces a \emph{fundamental tension}
between providing a strictly serialisable behaviour on the one hand,
versus availability and performance on the other.
This explains why large-scale data stores hardly ever provide the SSER
model, with the notable exception of Spanner \cite{rep:pan:1693}.

\subsection{Consistency models}


Informally, a \emph{consistency model} defines what an application can
observe about the updates and reads of its data store.
When the observed values of data differ from strict serialisability,
this is called an \emph{anomaly}.
Examples of anomalies include
\begin{inparablank}
\item
  \emph{divergence}, where concurrent reads of the same item
  persistently return different values;
\item
  \emph{causality violation}, where updates are observed out of order;
\item
  \emph{dirty reads}, where a read observes the effect of a
  transaction that has not terminated;
  or
\item
  \emph{lost updates}, where the effect of an update is lost.
\end{inparablank}
The more anomalies allowed by a store, the \emph{weaker} its consistency
model.
The \emph{strongest} model is (by definition) SSER, the baseline against
which other models are compared.

More formally, a consistency model is defined by the history of updates
and reads that clients can observe.
A model is weaker than another if it allows more histories.

An absolute definition of ``strong'' and ``weak consistency'' is open to
debate.
For the purpose of this entry, we say that a consistency model is
\emph{strong} when it has consensus power, i.e., any number of
failure-prone processes can reach agreement on some value by
communicating through data items in the store.
If this is not possible, then the consistency model is said \emph{weak}.

In a strong model, updates are totally ordered.
Some well-known strong models include SSER, serialisability (SER), or
snapshot isolation (SI).
Weak models admit concurrent updates to the same data item, and include
Causal Consistency (CC), Strong Eventual Consistency (SEC) and Eventual
Consistency (EC) (see Table~\ref{tab:sources}).



\section{Key Research Findings}

\subsection{Basic concepts}

A \emph{data store} is a logically-shared memory space where different application processes store,
update and retrieve data \emph{items}.
An item can be very basic, such as a register with read{\slash}write
operations, or a more complex structure, e.g., a table or a file system directory.
An application process executes \emph{operations} on the data store through the help of an \emph{API}.
When a process \emph{invokes} an operation, it executes a remote call to an appropriate \emph{end-point} of the data store.
In return, it receives a \emph{response value}.
A common example of this mechanism is a POST request to an HTTP end-point.


An application consists of \emph{transactions}.
A transaction consists of any number of reads and updates to the data store.
It is terminated either by an abort, whereby its writes have no effect,
or by a commit, whereby writes modify the store.
In what follows, we consider only committed transactions.
The transaction groups together low-level storage operations into a
higher-level abstraction, with properties that help developers reason
about application behaviour.

The properties of transactions are often summarised as ACID:
\underline{A}ll-or-Nothing, (individual) \underline{C}orrectness,
\underline{I}solation, and \underline{D}urability.

\emph{All-or-Nothing} ensures that, at any point in time, either all of a
transaction's writes are in the store, or none of them is.
%
%
This guarantee is essential in order to support common data invariants
such as equality or complementarity between two data items.
\emph{Individual Correctness} is the requirement that each of the
application's transactions individually transitions the database from a
safe state (i.e., where some application-specific integrity invariants
hold over the data) to another safe state.
\emph{Durability} means that all later transactions will observe the effect
of this transaction after it commits.
A, C and D are essential features of any transactional system, and will
be taken for granted in the rest of this entry.

The I property, \emph{Isolation} characterises the absence of
interference between transactions.
Transactions are isolated if one cannot interfere with the other,
regardless of whether they execute in parallel or not.

Taken together, the ACID properties provide the \emph{serialisability}
model, a program semantics in which a transaction executes as if it was
the only one accessing the database.%
%
%
This model restricts the allowable interactions among concurrent
transactions, such that each one produces results indistinguishable from
the same transactions running one after the other.
As a consequence, the code for a transaction lives
in the simple, familiar sequential programming world, and the developer 
can reason only about computations that start with the final results of
other transactions.
Serialisability allows concurrent operations to access the data store 
and still produce predictable, reproducible results.

\subsection{Definitions}

A \emph{history} is a sequence of invocations and responses of operations on the data items by the application processes.
It is commonly represented with timelines.
For instance, in history $h_1$ below, processes $p$ and $q$ access a data store that contains two counters ($x$ and $y$).
\vspace{-.5em}%
\tikzstyle{transA} = [color=blue]
\tikzstyle{transB} = [color=red]
\tikzstyle{transC} = [color=OliveGreen]

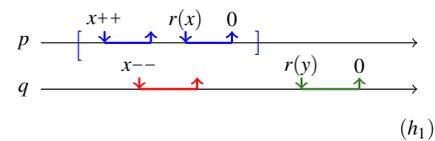
\tikzstyle{call} = [color=black,thick]
\tikzstyle{callA} = [color=blue,thick]
\tikzstyle{callB} = [color=red,thick]
\tikzstyle{callC} = [color=OliveGreen,thick]

\tikzset{
    ncbar angle/.initial=90,
    ncbar/.style={
        to path=(\tikztostart)
        -- ($(\tikztostart)!#1!\pgfkeysvalueof{/tikz/ncbar angle}:(\tikztotarget)$)
        -- ($(\tikztotarget)!($(\tikztostart)!#1!\pgfkeysvalueof{/tikz/ncbar angle}:(\tikztotarget)$)!\pgfkeysvalueof{/tikz/ncbar angle}:(\tikztostart)$)
        -- (\tikztotarget)
    },
    ncbar/.default=0.5cm,
}
\tikzset{square left brace/.style={ncbar=0.05cm}}
\tikzset{square right brace/.style={ncbar=-0.05cm}}

\begin{figure}[!h]
  \centering
  \fontsize{8}{11}\selectfont
  \begin{tikzpicture}[scale=0.77]
    \node at (7,.3) {$(h_1)$};
    
    \node at (0.2,1.8) {$p$};
    \node at (0.2,1) {$q$};
    
    \path[->] (0.5,1) edge (7,1);
    \path[->] (0.5,1.8) edge (7,1.8);
    
    \draw [transA] (1.2,1.5) to [square left brace ] (1.2,2);
    \path[callA] (1.6,1.8) edge (2.4,1.8);
    \path[callA,->] (1.6,2) edge (1.6,1.8);
    \path[callA,->] (2.4,1.8) edge (2.4,2);
    \node at (1.6,2.2) {$\inc{x}$};
    \node at (2.8,2.2) {};

    \path[callA] (3,1.8) edge (3.8,1.8);
    \path[callA,->] (3,2) edge (3,1.8);
    \path[callA,->] (3.8,1.8) edge (3.8,2);
    \node at (3,2.2) {$r(x)$};
    \node at (3.8,2.2) {$0$};
    \draw [transA] (4.2,1.6) to [square right brace ] (4.2,2);

    \path[callB] (2.2,1) edge (3.2,1);
    \path[callB,->] (2.2,1.2) edge (2.2,1);
    \path[callB,->] (3.2,1) edge (3.2,1.2);
    \node at (2.2,1.4) {$\dec{x}$};
    \node at (3.2,1.4) {};

    \path[callC] (5,1) edge (6,1);
    \path[callC,->] (5,1.2) edge (5,1);
    \path[callC,->] (6,1) edge (6,1.2);
    \node at (5,1.4) {$r(y)$};
    \node at (6,1.4) {$0$};
    
    \pgfresetboundingbox
    \clip[use as bounding box] (0,.3) rectangle (8,2.5);
  \end{tikzpicture}
\end{figure}

Operations $(\inc{z})$ and $(\dec{z})$ respectively increment and decrement counter $z$ by $1$.
To fetch the content of $z$, a process calls $r(z)$.
A counter is initialised to $0$.
The start and the end of a transaction are marked using brackets, e.g., transaction ${\color{blue}{T_1}}=(\inc{x}).r(x)$ in history $h_1$.
When the transaction contains a single operation, the brackets are omitted for clarity.

As pointed above, we assume in this entry that all the transactions are committed in a history.
Thus every invocation has a matching response.
More involved models exist, e.g., when considering a transactional memory \cite{Guerraoui:2008}.

A history induces a real-time order between transactions (denoted $\prec_h$).
This order holds between two transactions $T$ and $T'$ when the response of the last operation in $T$ precedes in $h$ the invocation of the first operation in $T'$.
A history also induces a per-process order that corresponds to the order in which processes invoke their transactions.
For instance in $h_1$, transaction ${\color{red}{T_2}}=(\dec{x})$ precedes transaction ${\color{OliveGreen}{T_3}}=r(y)$ at process $q$.
This relation together with (${\color{blue}{T_1}} <_{h_1} {\color{OliveGreen}{T_3}}$) fully defines the real-time order in history $h_1$.

Histories have various properties according to the way invocations and responses interleave.
Two transactions are concurrent in a history $h$ when they are not ordered by the relation $\prec_h$.
A history $h$ is \emph{sequential} when no two transactions in $h$ are concurrent.
A sequential history is \emph{legal} when it respects the sequential specification of each object.
Two histories $h$ and $h'$ are \emph{equivalent} when they contain the same set of events (invocations and responses).

A consistency model defines the histories that are allowed by the data store.
In particular, serialisability (SER) requires that every history $h$ is equivalent to some sequential and legal history $l$.
For instance, history $h_1$ is serialisable, since it is equivalent to the history $l_1$ below.
\vspace{-.5em}%
\begin{figure}[!h]
  \centering
  \fontsize{8}{11}\selectfont
  \begin{tikzpicture}[scale=0.77]
    \node at (7,.3) {$(l_1)$};

    \node at (0.2,1.8) {$p$};
    \node at (0.2,1) {$q$};
    
    \path[->] (0.5,1) edge (7,1);
    \path[->] (0.5,1.8) edge (7,1.8);
    
    \path[callB] (1,1) edge (2,1);
    \path[callB,->] (1,1.2) edge (1,1);
    \path[callB,->] (2,1) edge (2,1.2);
    \node at (1,1.4) {$\dec{x}$};
    \node at (2,1.4) {};
    
    \draw [transA] (2.2,1.5) to [square left brace ] (2.2,2);
    \path[callA] (2.6,1.8) edge (3.4,1.8);
    \path[callA,->] (2.6,2) edge (2.6,1.8);
    \path[callA,->]   (3.4,1.8) edge (3.4,2);
    \node at (2.6,2.2) {$\inc{x}$};
    \node at (3.6,2.2) {};

    \path[callA] (3.8,1.8) edge (4.6,1.8);
    \path[callA,->] (3.8,2) edge (3.8,1.8);
    \path[callA,->]   (4.6,1.8) edge (4.6,2);
    \node at (3.8,2.2) {$r(x)$};
    \node at (4.6,2.2) {$0$};
    \draw [transA] (5,1.6) to [square right brace ] (5,2);
    
    \path[callC] (5.6,1) edge (6.6,1);
    \path[callC,->] (5.6,1.2) edge (5.6,1);
    \path[callC,->] (6.6,1) edge (6.6,1.2);
    \node at (5.6,1.4) {$r(y)$};
    \node at (6.6,1.4) {$0$};

    \pgfresetboundingbox
    \clip[use as bounding box] (0,.3) rectangle (8,2.2);
    
  \end{tikzpicture}
\end{figure}
%
In addition, if the equivalent sequential history preserves the real-time order between transaction, history $h$ is said \emph{strictly serialisable} (SSER) \cite{Papadimitriou:1979}.
This is the case of $h_1$ since in $l_1$ the relations (${\color{red}{T_2}} <_{h_1} {\color{OliveGreen}{T_3}}$) and (${\color{blue}{T_1}} <_{h_1} {\color{OliveGreen}{T_3}}$) also hold.

When each transaction contains a single operation, SSER boils down to linearizability (LIN) \cite{loo:syn:1468}.
The data store ensures sequential consistency (SC) \cite{Lamport:1979}
when each transaction contains a single operation and only the
per-process order is kept in the equivalent sequential history.

The above consistency models (SER, SSER, LIN and SC) are strong, as they
allow the client application processes to reach consensus.
To see this, observe that processes may agree as follows:
The processes share a FIFO queue $L$ in the data store.
To reach consensus, each process enqueues some value in $L$ which corresponds to a proposal to agree upon.
Then, each process chooses the first proposal that appears in $L$.
The equivalence with a sequential history implies that all the application processes pick the same value.

Conversely, processes cannot reach consensus if the consistency model is \emph{weak}.
A widespread model in this category is Eventual Consistency (EC)
\cite{rep:syn:pan:1624}, used for instance in the Simple Storage Service
\cite{Murty:2008}.
EC requires that, if clients cease submitting transactions, they eventually observe the same state of the data store.
This eventually-stable state may include part (or all) the transactions
executed by the clients.
Under EC, processes may repeatedly observe updates in different orders.
For example, if the above list $L$ is EC, each process may see its
update applied first on $L$ until it decides, preventing agreement.
In fact, EC is too weak to allow asynchronous failure-prone processes to reach an agreement \cite{syn:rep:1738}.

\subsection{Fundamental results}

In the most general model of computation, replicas are asynchronous.
In this model, and under the hypothesis that a majority of them are correct,
it is possible to emulate a linearizable shared memory \cite{alg:mem:903}.
This number of correct replicas is tight.
In particular, if any majority of the replicas may fail, the emulation does not work \cite{Delporte-Gallet:2004}.

The above result implies that, even for a very basic distributed service, such as a register,
it is not possible to be at the same time consistent, available and tolerant to partition.
This result is known as the CAP Theorem \cite{rep:pan:1628}, which proves that it is not
possible to provide all the following desirable features at the same time:
\begin{inparaenum}[]
\item \emph{(C)} strong \underline{C}onsistency, even for a register,
%
%
\item \emph{(A)} \underline{A}vailability, responding to every client request, and 
\item \emph{(P)} tolerate network \underline{P}artition or arbitrary messages loss. 
\end{inparaenum}

A second fundamental result, known as FLP, is the impossibility to reach consensus
deterministically in presence of crash failures \cite{con:pan:640}.
FLP is true even if all the processes but one are correct.

As pointed above, a majority of correct processes may emulate a shared memory.
Thus, the FLP impossibility result indicates that a shared memory is not sufficient to reach consensus.
In fact, solving consensus requires the additional ability to elect a leader among the correct processes \cite{Chandra:1996}.

Data stores that support transactions on more than one data item are
subject to additional impossibility results.
For instance, an appealing property is genuine partial replication (GPR)
\cite{psutra-srds10}, a form of disjoint-access parallelism
\cite{Israeli:1994}.
Under GPR, transactions that access disjoint items do not contend in the data store.
GPR avoids convoy effects between transactions \cite{blasgen1979convoy} and ensure scalability under parallel workload.
However, GPR data stores must sacrifice some form of consistency, or provide little progress guarantees \cite{Bushkov:2014,psutra-europar13,Attiya:2009}.

A data store API defines the shared data structures the client application processes manipulate as well as their consistency and progress guarantees.
The above impossibility results inform the application developer that some APIs require synchronisation among the data replicas.
Process synchronisation is costly, thus there is a trade-off between performance and data consistency.

\subsection{{Trade-offs}}

In the common case, executing an operation under strong consistency requires to solve consensus among the data replicas, which costs at least one round-trip among replicas \cite{Lamport:2006}.
Sequential consistency allows to execute either read or write operations at a local replica \cite{Attiya:1994,Welch:2014}.
Weaker consistency models, e.g., eventual \cite{alg:rep:syn:optim:1464} and strong eventual consistency \cite{syn:rep:sh143} enable both read and write operations to be local.

A second category of trade-offs relate consistency models to metadata \cite{Peluso:2015,BurckhardtGY:Z014}.
They establish lower bounds on the space complexity to meet a certain consitency models.
For instance, tracking causality accurately requires $O(m)$ bits of storage, where $m$ is the number of replicas \cite{charronBost:1991}.

\begin{table*}[t]
  \centering
    \begin{tabular}[t]{@{}ccc@{}}
      Acronym & Full name & Reference \\
      \hline
      \citeacronym{EC}
      \citeacronym{SEC}
      \citeacronym{CM}
      \citeacronym{CS}
      \citeacronym{CC}
      \citeacronym{Causal HAT}
      \citeacronym{LIN}
      \citeacronym{NMSI}
      \citeacronym{PSI}
      \citeacronym{RC}
      \citeacronym{SC}
      \citeacronym{SER}
      \citeacronym{SI}
      \citeacronym{SSER}
      \citeacronym{SSI}
    \end{tabular}
  \caption{Models and source references}
  \label{tab:sources}
\end{table*}

\subsection{Common models}

The previous sections introduce several consistency models (namely, SER, SC, LIN, SSER and EC).
This section offers a perspective on other prominent models.
Table~\ref{tab:sources} recapitulates.

\paragraph{Read-Committed (RC).}
Almost all existing transactional data stores ensure that clients observe only committed data \cite{Zemke:2012,syn:bd:1759}.
More precisely, the RC consistency model enforces that if some read $r$ observes the state $\hat{x}$ of an item $x$ in history $h$, then the transaction $T_i$ that wrote $\hat{x}$ commits in $h$.
One can distinguish a \emph{loose} and a \emph{strict} interpretation of RC.
The strict interpretation requires that $r(x)$ takes place after transaction $T_i$ commits.
Under the loose interpretation, the write operation might occur concurrently.

When RC, or a stricter consistency model holds, it is convenient to introduce the notion of \emph{version}.
A version is the state of a data item as produced by an update transaction.
For instance, when $T_i$ writes to some register $x$, an operation denoted hereafter $w(x_i)$, it creates a new version $x_i$ of $x$.
Versions allow to uniquely identify the state of the item as observed by a read operation, e.g., $r(x_i)$.

\paragraph{Strong Eventual Consistency (SEC).}
Eventual consistency (EC) states that, for every data item $x$ in the store, if there is no new update on $x$, eventually clients observe $x$ in the same state.
Strong eventual consistency (SEC) further constrains the behaviour of the data replicas.
In detail, a data store is SEC when it is EC and moreover, for every item $x$, any two replicas of $x$ that applied the same set of updates on item $x$ are in the same state.

\paragraph{Client Monotonic (CM).}
Client Monotonic (CM) ensures that a client always observes the results of its own past operations \cite{rep:syn:1481}.
CM enforces the following four so-called ``session guarantees'':
\begin{inparaenum}[\em(i)]
\item Monotonic reads (MR): if a client executes $r(x_i)$ then $r(x_{j  \neq i})$ in history $h$, necessarily $x_j$ follows $x_i$ for some version order $\ll_{h,x}$ over the updates applied to $x$ in $h$.
\item Monotonic writes (MW): if a client executes $w(x_i)$ then $w(x_j)$, the version order $x_i \ll_{h,x} x_j$ holds;
\item Read-my-writes (RMW): when a client executes $w(x_i)$ followed by $r(x_{j \neq i})$, then $x_i \ll_{h,x} x_j$ holds; and
\item Writes-follow-reads (WFR): if a client executes $r(x_i)$ followed by $w(x_j)$ it is true that $x_i \ll_{h,x} x_j$.
\end{inparaenum}


Most consistency models require CM, but this guarantee is so obvious that it might be sometimes omitted -- this is for instance the case in \citet{Gray:1992}.

\paragraph{Read-Atomic (RA).}
Under RA, a transaction sees either all of the updates made by
another transaction, or none of them (the All-or-Nothing guarantee).
For instance, if a transaction $T$ sees the version $x_i$ written by $T_i$
and transaction $T_i$ also updates $y$, then $T$ should observe at least version $y_i$.
If history $h$ fails to satisfies RA, a transaction in $h$ exhibits a \emph{fractured read} \cite{Bailis:2014}.
For instance, this is the case of the transaction executed by process $q$ in history $h_2$ below.
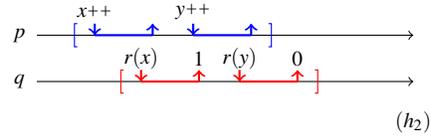
\vspace{-.5em}%
\begin{figure}[!h]
  \centering
  \fontsize{8}{11}\selectfont
  \begin{tikzpicture}[scale=0.77]
    \node at (7,.3) {$(h_2)$};
    
    \node at (0.2,1.8) {$p$};
    \node at (0.2,1) {$q$};
    
    \path[->] (0.5,1) edge (7,1);
    \path[->] (0.5,1.8) edge (7,1.8);

    \draw [transA] (1.2,1.6) to [square left brace ] (1.2,2);
    \path[callA] (1.5,1.8) edge (2.5,1.8);
    \path[callA,->] (1.5,2) edge (1.5,1.8);
    \path[callA,->]   (2.5,1.8) edge (2.5,2);
    \node at (1.5,2.2) {$\inc{x}$};
    \node at (2.5,2.2) {};

    \path[callA] (3.2,1.8) edge (4.2,1.8);
    \path[callA,->] (3.2,2) edge (3.2,1.8);
    \path[callA,->]   (4.2,1.8) edge (4.2,2);
    \node at (3.2,2.2) {$\inc{y}$};
    \node at (4.2,2.2) {};
    \draw [transA] (4.5,1.6) to [square right brace ] (4.5,2);

    \draw [transB] (2,.8) to [square left brace ] (2,1.2);
    \path[callB] (2.3,1) edge (3.3,1);
    \path[callB,->] (2.3,1.2) edge (2.3,1);
    \path[callB,->]   (3.3,1) edge (3.3,1.2);
    \node at (2.3,1.4) {$r(x)$};
    \node at (3.3,1.4) {$1$};

    \path[callB] (4,1) edge (5,1);
    \path[callB,->] (4,1.2) edge (4,1);
    \path[callB,->]   (5,1) edge (5,1.2);
    \node at (4,1.4) {$r(y)$};
    \node at (5,1.4) {$0$};
    \draw [transB] (5.3,.8) to [square right brace ] (5.3,1.2);

    \pgfresetboundingbox
    \clip[use as bounding box] (0,.3) rectangle (8,2.5);
  \end{tikzpicture}
\end{figure}
\vspace{-1em}%

\paragraph{Consistent Snapshot (CS).}
A transaction $T_i$ \emph{depends on} a transaction $T_j$ when it reads a version written by $T_j$, or such a relation holds transitively.
In other words, denoting $T_i \readFrom T_j$ when $T_i$ reads from $T_j$,
$T_j$ is in the transitive closure of the relation $(\readFrom)$ when starting from $T_i$.

When a transaction never misses the effects of some transaction it depends on, the transaction observes a \emph{consistent snapshot} \cite{Chan:1985}.
In more formal terms, a transaction $T_i$ in a history $h$ observes a consistent snapshot when for every object $x$, if
\begin{inparaenum}[\em(i)]
\item $T_i$ reads version $x_j$,
\item $T_k$ writes version $x_k$, and 
\item $T_i$ depends on $T_{k}$,
\end{inparaenum}
then version $x_k$ is followed by version $x_j$ in the version order $\ll_{h,x}$.
A history $h$ belongs to CS when all its transactions observe a consistent snapshot.
For instance, this is not the case of history $h_3$ below.
In this history, transaction ${\color{OliveGreen}{T_3}}=r(y).r(x)$ depends on ${\color{red}{T_2}}=r(x).(\inc{y})$, and $T_2$ depends on ${\color{blue}{T_1}}=(\inc{x})$, yet $\color{OliveGreen}{T_3}$ does not observe the effect of $\color{blue}{T_1}$.
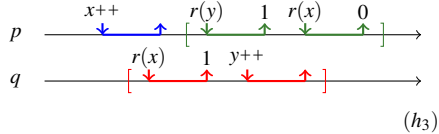
\vspace{-.5em}%
\begin{figure}[!h]
  \centering
  \fontsize{8}{11}\selectfont
  \begin{tikzpicture}[scale=0.77]
    \node at (7,.3) {$(h_3)$};
    
    \node at (0,1.8) {$p$};
    \node at (0,1) {$q$};
    
    \path[->] (0.5,1) edge (7,1);
    \path[->] (0.5,1.8) edge (7,1.8);

    \path[callA] (1.5,1.8) edge (2.5,1.8);
    \path[callA,->] (1.5,2) edge (1.5,1.8);
    \path[callA,->]   (2.5,1.8) edge (2.5,2);
    \node at (1.5,2.2) {$\inc{x}$};
    \node at (2.5,2.2) {};

    \draw [transB] (2,.8) to [square left brace ] (2,1.2);
    \path[callB] (2.3,1) edge (3.3,1);
    \path[callB,->] (2.3,1.2) edge (2.3,1);
    \path[callB,->]   (3.3,1) edge (3.3,1.2);
    \node at (2.3,1.4) {$r(x)$};
    \node at (3.3,1.4) {$1$};

    \path[callB] (4,1) edge (5,1);
    \path[callB,->] (4,1.2) edge (4,1);
    \path[callB,->]   (5,1) edge (5,1.2);
    \node at (4,1.4) {$\inc{y}$};
    \node at (5,1.4) {};
    \draw [transB] (5.3,.8) to [square right brace ] (5.3,1.2);

    \draw [transC] (3,1.6) to [square left brace ] (3,2);
    \path[callC] (3.3,1.8) edge (4.3,1.8);
    \path[callC,->] (3.3,2) edge (3.3,1.8);
    \path[callC,->]   (4.3,1.8) edge (4.3,2);
    \node at (3.3,2.2) {$r(y)$};
    \node at (4.3,2.2) {$1$};

    \path[callC] (5,1.8) edge (6,1.8);
    \path[callC,->] (5,2) edge (5,1.8);
    \path[callC,->]   (6,1.8) edge (6,2);
    \node at (5,2.2) {$r(x)$};
    \node at (6,2.2) {$0$};
    \draw [transC] (6.3,1.6) to [square right brace ] (6.3,2);
   
    \pgfresetboundingbox
    \clip[use as bounding box] (0,.3) rectangle (8,2.5);
  \end{tikzpicture}
\end{figure}
\paragraph{Causal Consistency (CC).}
Causal consistency (CC) holds when transactions observe consistent snapshots of the system, and the client application processes are monotonic.
CC is a weak consistency model and it does not allow solving  consensus.
It is in fact the strongest model that is available under partition \cite{syn:rep:1738}.
Historically, CC refers to the consistency of single operations on a shared memory \cite{syn:mat:1025}.
Causally consistent transactions (Causal HAT) is a consistency model that extends CC to transactional data stores \cite{db:syn:1751}.

\paragraph{Snapshot Isolation (SI).}
SI is a widely-used consistency model \cite{syn:bd:1759}.
This model is strong, but allows more interleavings of concurrent read transactions than SER.
Furthermore, SI is causal (i.e., SI $\subseteq$ CC), whereas SER is not.

Under SI, a transaction observes a \emph{snapshot} of the state of the data store at some point prior in time.
Strong snapshot isolation (SSI) requires this snapshot to contain all
the preceding transactions in real time \cite{Daudjee:2006}.
Two transactions may commit under SI as long as they do not write the same item concurrent.
SI avoids the  anomalies listed in \secref{sec:consistency-models}, but
exhibits the \emph{write-skew} anomaly, illustrated in history $h_4$ below.
\vspace{-.5em}%
\begin{figure}[!h]
  \centering
  \fontsize{8}{11}\selectfont
  \begin{tikzpicture}[scale=0.77]
    \node at (7,.3) {$(h_4)$};
    
    \node at (0,1.8) {$p$};
    \node at (0,1) {$q$};
    
    \path[->] (0.5,1) edge (7,1);
    \path[->] (0.5,1.8) edge (7,1.8);

    \draw [transA] (.8,1.6) to [square left brace ] (.8,2);
    \path[callA] (1.1,1.8) edge (2.1,1.8);
    \path[callA,->] (1.1,2) edge (1.1,1.8);
    \path[callA,->]   (2.1,1.8) edge (2.1,2);
    \node at (1.1,2.2) {$r(x)$};
    \node at (2.1,2.2) {$2$};

    \path[callA] (2.6,1.8) edge (3.8,1.8);
    \path[callA,->] (2.8,2) edge (2.8,1.8);
    \path[callA,->]   (3.8,1.8) edge (3.8,2);
    \node at (2.8,2.2) {$r(y)$};
    \node at (3.8,2.2) {$1$};

    \path[callA] (4.4,1.8) edge (5.4,1.8);
    \path[callA,->] (4.4,2) edge (4.4,1.8);
    \path[callA,->]   (5.4,1.8) edge (5.4,2);
    \node at (4.4,2.2) {$\dec{x}$};
    \node at (5.4,2.2) {};
    \draw [transA] (5.7,1.6) to [square right brace ] (5.7,2);

    \draw [transB] (.8,.8) to [square left brace ] (.8,1.2);
    \path[callB] (1.1,1) edge (2.1,1);
    \path[callB,->] (1.1,1.2) edge (1.1,1);
    \path[callB,->]   (2.1,1) edge (2.1,1.2);
    \node at (1.1,1.4) {$r(x)$};
    \node at (2.1,1.4) {$2$};

    \path[callB] (2.8,1) edge (3.8,1);
    \path[callB,->] (2.8,1.2) edge (2.8,1);
    \path[callB,->]   (3.8,1) edge (3.8,1.2);
    \node at (2.8,1.4) {$r(y)$};
    \node at (3.8,1.4) {$1$};

    \path[callB] (4.4,1) edge (5.4,1);
    \path[callB,->] (4.4,1.2) edge (4.4,1);
    \path[callB,->]   (5.4,1) edge (5.4,1.2);
    \node at (4.4,1.4) {$\inc{y}$};
    \node at (5.4,1.4) {};
    \draw [transB] (5.8,.8) to [square right brace ] (5.8,1.2);
    
    \pgfresetboundingbox
    \clip[use as bounding box] (0,.3) rectangle (8,2.5);
  \end{tikzpicture}
\end{figure}
In this history, an application using data items $x$ and $y$ wishes to maintain the invariant $x \geq y$.
The invariant holds initially, and each of the two transactions
${\color{blue}{T_1}}$ and ${\color{red}{T_2}}$ guarantees the invariant
individually.
As illustrated in history $h_4$, running them concurrently under SI may violate the invariant.

An application is \emph{robust} against a consistency model $M$ when,
it produces serialisable histories \cite{Cerone:2016},
despite running atop a data store providing $M$, 
It is known \cite{syn:1687} that an application is robust against SI when every
invariant is materialised by a data item.

\paragraph{Parallel {\slash} Non-Monotonic Snapshot Isolation (PSI{\slash}NMSI).}
Parallel and non-monotonic snapshot isolation are scalable variations of SI.
These models retain two core properties of SI, namely
\begin{inparaenum}[\em(i)]
\item each transaction observes a consistent snapshot, and
\item no two concurrent transactions update the same data items.
\end{inparaenum}
PSI requires to take a snapshot at the start of the transaction.
NMSI relaxes this requirement, enabling the snapshot to be computed
incrementally, as illustrated in history $h_5$ below.
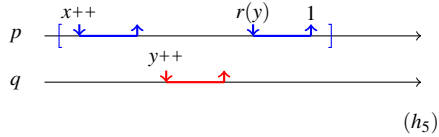
\vspace{-.5em}%
\begin{figure}[!h]
  \centering
  \fontsize{8}{11}\selectfont
  \begin{tikzpicture}[scale=0.77]
    \node at (7,.3) {$(h_5)$};
    
    \node at (0,1.8) {$p$};
    \node at (0,1) {$q$};
    
    \path[->] (0.5,1) edge (7,1);
    \path[->] (0.5,1.8) edge (7,1.8);

    \draw [transA] (.8,1.6) to [square left brace ] (.8,2);
    \path[callA] (1.1,1.8) edge (2.1,1.8);
    \path[callA,->] (1.1,2) edge (1.1,1.8);
    \path[callA,->]   (2.1,1.8) edge (2.1,2);
    \node at (1.1,2.2) {$\inc{x}$};
    \node at (2.1,2.2) {};

    \path[callA] (4.1,1.8) edge (5.1,1.8);
    \path[callA,->] (4.1,2) edge (4.1,1.8);
    \path[callA,->]   (5.1,1.8) edge (5.1,2);
    \node at (4.1,2.2) {$r(y)$};
    \node at (5.1,2.2) {$1$};
    \draw [transA] (5.4,1.6) to [square right brace ] (5.4,2);
    
    \path[callB] (2.6,1) edge (3.6,1);
    \path[callB,->] (2.6,1.2) edge (2.6,1);
    \path[callB,->] (3.6,1) edge (3.6,1.2);
    \node at (2.6,1.4) {$\inc{y}$};
    \node at (3.6,1.4) {};

    \pgfresetboundingbox
    \clip[use as bounding box] (0,.3) rectangle (8,2.5);
  \end{tikzpicture}
\end{figure}
\subsection{A three-dimensional view of data consistency}

\citet{syn:formel:sh190} classify consistency models along three dimensions, to better understand and compare them.
Their approach divides each operation into two parts:
the \emph{generator} reads data and computes response values, and the \emph{effector} applies side-effects to every replica.
Each of the three dimensions imposes constraints on the generators and
effectors.
Table~\ref{tab:crossref} classifies the consistency criteria of Table~\ref{tab:sources} along these three dimensions.
\begin{itemize}
\item \emph{Visibility dimension.}
  This dimension constrains the visibility of operations, i.e., how a generator sees the updates made by effectors.
  The strongest class of consistency models along this dimension is \emph{external} visibility,
  which imposes that a generator sees the effectors of all the operations that precedes it in real time.
  Weakening this guarantee to the per-process order leads to \emph{causal} visibility.
  A yet weaker class is \emph{transitive} visibility, which only requires visibility to hold transitively.
  Finally, absence of constraints on generators, for instance during the unstable period of an eventually-consistent data store, is termed \emph{rollback} visibility.
  
\item \emph{Ordering dimension.}
  This dimension constrains the ordering over generators and effectors.
  Four classes are of interest along this dimension:
  The strongest class is termed \emph{total} order.
  For every history of a model in this class, there exists an equivalent serial history of all the operations.
  Weaker classes, below total order, constrain only effectors.
  The \emph{gapless} order class requires effectors to be ordered online by natural numbers with no gaps;
  this requires consensus and is subject to the CAP impossibility result.
  The \emph{capricious} class admits gaps in the ordering, allowing replicas to order their operations independently.
  A last-writer wins protocol (e.g., \cite{rep:alg:717bis}) produces a consistency model in this class.
  This class is subject to the lost-update anomaly.
  The weakest class along this dimension is termed \emph{concurrent} and imposes no ordering on generators and effectors.
  
\item \emph{Composition dimension.}
  This dimension captures the fact that a transaction contains one or more operations.
  A model in the All-Or-Nothing class preserves the A in ACID.
  This means that if some effector of transaction $T_1$ is visible to transaction $T_2$, then all of $T_1$’s effectors are visible to $T_2$.
  Typically, all the generators of a transaction read from the same set of effectors, i.e., its snapshot.
  The snapshot class extends the Visibility and Ordering guarantees to all generators of the transaction.
  For instance, in the case of a model both in the snapshot and total order classes, all the operations of a transaction are adjacent in the equivalent serial history.  
\end{itemize}

\begin{table*}[t]
  \centering
    \begin{tabular}[t]{@{}c|c|c|c@{}}
      Acronym & Ordering & Visibility & Composition \\
      \hline
      \longentry{EC}
      \longentry{CM}
      \longentry{CS}
      \longentry{CC}
      \longentry{Causal HAT}
      \longentry{LIN}
      \longentry{NMSI}
      \longentry{PSI}
      \longentry{RC}
      \longentry{SC}
      \longentry{SER}
      \longentry{SI}
      \longentry{SSER}
      \longentry{SSI}
    \end{tabular}
  \caption{Three-dimension features of consistency models and systems}
  \label{tab:crossref}
\end{table*}

\section{Examples of Application}

A key-value store (KVS) is a distributed data store that serves as building block of many cloud applications.
This type of system belongs to the larger family of NoSQL databases and is used to store uninterpreted blobs of data (e.g., marshalled objects).

A KVS implements a map, that is a mutable relation from a set of \emph{keys} to a set of \emph{values}.
In detail, the API of a key-value store consists of two operations:
Operation $\aput(k,v)$ adds the pair $(k,v)$ to the mapping, updating it if necessary.
Depending on the KVS, this operation may return the previous value of the key $k$, or simply $\nil$.
Operation $\get(k)$ retrieves the current value stored under key $k$.

The notions of ``current'' and ``previous'' values depend on the consistency model of the KVS.
History $h_6$ below illustrates this point for an operation $\get(k_1)$ by process $p$.
\vspace{-.5em}%
\begin{figure}[!h]
  \centering
  \fontsize{8}{11}\selectfont
  \begin{tikzpicture}[scale=0.77]
    \node at (7,-.3) {$(h_6)$};
    
    \node at (0,1.8) {$p$};
    \node at (0,1) {$q$};
    \node at (0,.2) {$q'$};    

    \path[->] (0.5,1.8) edge (7,1.8);
    \path[->] (0.5,1) edge (7,1);
    \path[->] (0.5,.2) edge (7,.2);
    
    \path[callA] (1,1.8) edge (1.6,1.8);
    \path[callA,->] (1,2) edge (1,1.8);
    \path[callA,->]   (1.6,1.8) edge (1.6,2);
    \node at (1,2.2) {$\aput(k_1,x)$};

    \path[callB] (2,1) edge (2.7,1);
    \path[callB,->] (2,1.2) edge (2,1);
    \path[callB,->]   (2.7,1) edge (2.7,1.2);
    \node at (2,1.4) {$\aput(k_1,y)$};

    \path[callC] (4,1) edge (4.7,1);
    \path[callC,->] (4,1.2) edge (4,1);
    \path[callC,->]   (4.7,1) edge (4.7,1.2);
    \node at (4,1.4) {$\aput(k_1,z)$};
    
    \path[call] (2.3,.2) edge (3.3,.2);
    \path[call,->] (2.3,.4) edge (2.3,.2);
    \path[call,->]   (3.3,.2) edge (3.3,.4);
    \node at (2.3,.6) {$\get(k_1)$};
    \node at (3.3,.6) {$y$};
    
    \path[call] (4.5,.2) edge (6,.2);
    \path[call,->] (4.5,.4) edge (4.5,.2);
    \path[call,->]   (6,.2) edge (6,.4);
    \node at (4.5,.6) {$\aput(k_2,1)$};
    
    \path[call] (3,1.8) edge (4.8,1.8);
    \path[call,->] (3,2) edge (3,1.8);
    \path[call,->]   (4.8,1.8) edge (4.8,2);
    \node at (3,2.2) {$\get(k_2)$};
    \node at (4.8,2.2) {$1$};

    \path[call] (5.6,1.8) edge (6.6,1.8);
    \path[call,->] (5.6,2) edge (5.6,1.8);
    \path[call,->]   (6.6,1.8) edge (6.6,2);
    \node at (5.6,2.2) {$\get(k_1)$};
    \node at (6.6,2.2) {\textbf{?}};
    
    \pgfresetboundingbox
    \clip[use as bounding box] (0,-.3) rectangle (8,2.5);
  \end{tikzpicture}
\end{figure}
When the KVS is RC, operation $\get(k_1)$ returns the initial value of $k_1$, or any value written concurrently or before this call.
Denoting $\bot$ the initial value of $k_1$, this means that operation $\get(k_1)$ may return any value in $\{\bot,{\color{blue}{x}},{\color{red}{y}},{\color{OliveGreen}{z}}\}$.

If the KVS guarantees RMW, at least the last value written by $p$ should be returned.
As a consequence, the set of possible values reduces to $\{{\color{blue}{x}},{\color{red}{y}},{\color{OliveGreen}{z}}\}$.

Now, let us consider that the KVS guarantees CC.
Process $p$ observes the operation $\aput(k_2,1)$ by $r$.
This operation causally follows an observation by $q'$ of ${\color{red}{y}}$.
Therefore, $p$ should observe either ${\color{red}{y}}$, or ${\color{OliveGreen}{z}}$.

If the KVS is linearizable, the value stored under key $k_1$ is the last value written before $\get(k_1)$ in any sequential history equivalent to $h_6$.
Every such history should preserve the real-time precedence of $h_6$.
Clearly, the last update in $h_6$ sets the value of $k_1$ to ${\color{OliveGreen}{z}}$.
Thus, if the KVS is linearizable, ${\color{OliveGreen}{z}}$ is the only allowed response of operation $\get(k_1)$ in $h_6$.

\section{Future Directions for Research}

Consistency models are formulated in various frameworks and using different underlying assumptions.
For instance, some works \cite{sql92,syn:bd:1759} define a model in terms of the anomalies it forbids.
Others rely on specific graphs to characterise a model \cite{Adya99}, or predicates over histories \cite{Viotti:2016}.
The existence of a global time \cite{Papadimitriou:1979} is sometimes taken for granted.
This contrasts with approaches \cite{Lamport:1986} that avoid to make such an assumption.
A similar observation holds for concurrent operations which may \cite{Guerraoui:2008} or not \cite{Ozsu:1991} overlap in time.

This rich literature makes difficult an apples-to-apples comparison between consistency models.
Works exist \cite{Chrysanthis:1994} that attempt to bridge this gap by expressing them in a common framework.
However, not all the literature is covered and it is questionable whether their definitions
is equivalent to the ones given in the original publications.

The general problem of the implementability a given model is also an interesting avenue for research.
One may address this question in term of the minimum synchrony assumptions to attain a particular model.
In distributed systems, this approach has lead to the rich literature on failure detectors \cite{Freiling:2011}.
A related question is to establish lower and upper bounds on the time and space complexity of an implementation (when it is achivable).
As pointed out in \secref{sec:trade-offs}, some results already exist, yet the picture is incomplete.

From an application point of view, three questions are of particular interest.
First, the robustness of an application against a particular consistency model \cite{syn:1687,Cerone:2016}.
Second, the relation between a model and a consistency control protocol.
These two questions are related to the grand challenge of synthesising concurrency
control from the application specification \cite{Gotsman:2016}.
A third challenge is to compare consistency models in practice \cite{Kemme:1998,Wiesmann:2005,SaeidaArdekani:2014},
so as to understand their pros and cons.

\renewcommand{\bibname}{References}

\ifthenelse{\boolean{isSpringer}}{
  \bibliographystyle{spbasic} 
}{
  \bibliographystyle{plainnat}
}
\bibliography{predef,shapiro-bib-ext,shapiro-bib-perso,sutra-bib-perso,sutra-bib-ext}

\begin{thebibliography}{54}
\providecommand{\natexlab}[1]{#1}
\providecommand{\url}[1]{{#1}}
\providecommand{\urlprefix}{URL }
\expandafter\ifx\csname urlstyle\endcsname\relax
  \providecommand{\doi}[1]{DOI~\discretionary{}{}{}#1}\else
  \providecommand{\doi}{DOI~\discretionary{}{}{}\begingroup
  \urlstyle{rm}\Url}\fi
\providecommand{\eprint}[2][]{\url{#2}}

\bibitem[{Adya(1999)}]{Adya99}
Adya A (1999) {Weak Consistency: A Generalized Theory and Optimistic
  Implementations for Distributed Transactions}. Ph.d., MIT

\bibitem[{Ahamad et~al(1995)Ahamad, Neiger, Burns, Kohli, and
  Hutto}]{syn:mat:1025}
Ahamad M, Neiger G, Burns JE, Kohli P, Hutto PW (1995) Causal memory:
  definitions, implementation, and programming. Distributed Computing
  9(1):37--49

\bibitem[{{ANSI}(1992)}]{sql92}
{ANSI} (1992)  {American National Standard for Information Systems -- Database
  Language -- SQL}

\bibitem[{Attiya and Welch(1994)}]{Attiya:1994}
Attiya H, Welch JL (1994) Sequential consistency versus linearizability. ACM
  Trans Comput Syst 12(2):91--122

\bibitem[{Attiya et~al(1990)Attiya, Bar-Noy, and Dolev}]{alg:mem:903}
Attiya H, Bar-Noy A, Dolev D (1990) Sharing memory robustly in message-passing
  systems. Tech. Rep. MIT/LCS/TM-423, Massachusetts Institute of Technology,
  Lab.\ for Comp.\ Sc., Cambridge, MA {(USA)}

\bibitem[{Attiya et~al(2009)Attiya, Hillel, and Milani}]{Attiya:2009}
Attiya H, Hillel E, Milani A (2009) Inherent limitations on disjoint-access
  parallel implementations of transactional memory. In: Proceedings of the
  Twenty-first Annual Symposium on Parallelism in Algorithms and Architectures,
  ACM, New York, NY, USA, SPAA '09, pp 69--78

\bibitem[{Attiya et~al(2017)Attiya, Ellen, and Morrison}]{syn:rep:1738}
Attiya H, Ellen F, Morrison A (2017) Limitations of highly-available
  eventually-consistent data stores. IEEE Trans\ on Parallel and Dist\ Sys
  (TPDS) 28(1):141--155

\bibitem[{Bailis et~al(2013)Bailis, Davidson, Fekete, Ghodsi, Hellerstein, and
  Stoica}]{db:syn:1751}
Bailis P, Davidson A, Fekete A, Ghodsi A, Hellerstein JM, Stoica I (2013)
  Highly available transactions: Virtues and limitations. Proc {VLDB} {E}ndow
  7(3):181--192

\bibitem[{Bailis et~al(2014)Bailis, Fekete, Hellerstein, Ghodsi, and
  Stoica}]{Bailis:2014}
Bailis P, Fekete A, Hellerstein JM, Ghodsi A, Stoica I (2014) Scalable atomic
  visibility with ramp transactions. In: Proceedings of the 2014 ACM SIGMOD
  International Conference on Management of Data, ACM, New York, NY, USA,
  SIGMOD '14, pp 27--38

\bibitem[{Berenson et~al(1995)Berenson, Bernstein, Gray, Melton, O'Neil, and
  O'Neil}]{syn:bd:1759}
Berenson H, Bernstein P, Gray J, Melton J, O'Neil E, O'Neil P (1995) A critique
  of {ANSI} {SQL} isolation levels. SIGMOD Rec 24(2):1--10

\bibitem[{Blasgen et~al(1979)Blasgen, Gray, Mitoma, and
  Price}]{blasgen1979convoy}
Blasgen M, Gray J, Mitoma M, Price T (1979) {The convoy phenomenon}. ACM SIGOPS
  Operating Systems Review 13(2):20--25

\bibitem[{Burckhardt et~al(2014)Burckhardt, Gotsman, Yang, and
  Zawirski}]{BurckhardtGY:Z014}
Burckhardt S, Gotsman A, Yang H, Zawirski M (2014) Replicated data types:
  specification, verification, optimality. In: The 41st Annual {ACM}
  {SIGPLAN-SIGACT} Symposium on Principles of Programming Languages, {POPL}
  '14, San Diego, CA, USA, January 20-21, 2014, pp 271--284

\bibitem[{Bushkov et~al(2014)Bushkov, Dziuma, Fatourou, and
  Guerraoui}]{Bushkov:2014}
Bushkov V, Dziuma D, Fatourou P, Guerraoui R (2014) The pcl theorem:
  Transactions cannot be parallel, consistent and live. In: Proceedings of the
  26th ACM Symposium on Parallelism in Algorithms and Architectures, ACM, New
  York, NY, USA, SPAA '14, pp 178--187

\bibitem[{Cerone and Gotsman(2016)}]{Cerone:2016}
Cerone A, Gotsman A (2016) Analysing snapshot isolation. In: Proceedings of the
  2016 {ACM} Symposium on Principles of Distributed Computing, {PODC} 2016,
  Chicago, IL, USA, July 25-28, 2016, pp 55--64

\bibitem[{Chan and Gray(1985)}]{Chan:1985}
Chan A, Gray R (1985) {Implementing Distributed Read-Only Transactions}. IEEE
  Trans Softw Eng SE-11(2):205--212

\bibitem[{Chandra et~al(1996)Chandra, Hadzilacos, and Toueg}]{Chandra:1996}
Chandra TD, Hadzilacos V, Toueg S (1996) The weakest failure detector for
  solving consensus. J ACM 43(4):685--722

\bibitem[{Charron{-}Bost(1991)}]{charronBost:1991}
Charron{-}Bost B (1991) Concerning the size of logical clocks in distributed
  systems. Inf Process Lett 39(1):11--16

\bibitem[{Chrysanthis and Ramamritham(1994)}]{Chrysanthis:1994}
Chrysanthis PK, Ramamritham K (1994) {Synthesis of Extended Transaction Models
  Using ACTA}. ACM Trans Database Syst 19(3):450--491

\bibitem[{Corbett et~al(2012)Corbett, Dean, Epstein, Fikes, Frost, Furman,
  Ghemawat, Gubarev, Heiser, Hochschild, Hsieh, Kanthak, Kogan, Li, Lloyd,
  Melnik, Mwaura, Nagle, Quinlan, Rao, Rolig, Saito, Szymaniak, Taylor, Wang,
  and Woodford}]{rep:pan:1693}
Corbett JC, Dean J, Epstein M, Fikes A, Frost C, Furman J, Ghemawat S, Gubarev
  A, Heiser C, Hochschild P, Hsieh W, Kanthak S, Kogan E, Li H, Lloyd A, Melnik
  S, Mwaura D, Nagle D, Quinlan S, Rao R, Rolig L, Saito Y, Szymaniak M, Taylor
  C, Wang R, Woodford D (2012) {S}panner: {G}oogle's globally-distributed
  database. In: Symp.\ on Op.\ Sys.\ Design and Implementation (OSDI), Usenix,
  Hollywood, CA, USA, pp 251--264

\bibitem[{Daudjee and Salem(2006)}]{Daudjee:2006}
Daudjee K, Salem K (2006) Lazy database replication with snapshot isolation.
  In: Proceedings of the 32Nd International Conference on Very Large Data
  Bases, VLDB Endowment, VLDB '06, pp 715--726

\bibitem[{Delporte-Gallet et~al(2004)Delporte-Gallet, Fauconnier, Guerraoui,
  Hadzilacos, Kouznetsov, and Toueg}]{Delporte-Gallet:2004}
Delporte-Gallet C, Fauconnier H, Guerraoui R, Hadzilacos V, Kouznetsov P, Toueg
  S (2004) The weakest failure detectors to solve certain fundamental problems
  in distributed computing. In: Proceedings of the Twenty-third Annual ACM
  Symposium on Principles of Distributed Computing, ACM, New York, NY, USA,
  PODC '04, pp 338--346

\bibitem[{Fekete et~al(1999)Fekete, Gupta, Luchangco, Lynch, and
  Shvartsman}]{alg:rep:syn:optim:1464}
Fekete A, Gupta D, Luchangco V, Lynch N, Shvartsman A (1999)
  Eventually-serializable data services. Theoretical Computer Science
  220:113--156, special issue on Distributed Algorithms

\bibitem[{Fekete et~al(2005)Fekete, Liarokapis, O'Neil, O'Neil, and
  Shasha}]{syn:1687}
Fekete A, Liarokapis D, O'Neil E, O'Neil P, Shasha D (2005) Making snapshot
  isolation serializable. Trans\ on Database Systems 30(2):492--528

\bibitem[{Fischer et~al(1985)Fischer, Lynch, and Patterson}]{con:pan:640}
Fischer MJ, Lynch NA, Patterson MS (1985) Impossibility of distributed
  consensus with one faulty process. J~ACM 32(2):374--382

\bibitem[{Freiling et~al(2011)Freiling, Guerraoui, and
  Kuznetsov}]{Freiling:2011}
Freiling FC, Guerraoui R, Kuznetsov P (2011) The failure detector abstraction.
  ACM Comput Surv 43(2):9:1--9:40

\bibitem[{Gilbert and Lynch(2002)}]{rep:pan:1628}
Gilbert S, Lynch N (2002) Brewer's conjecture and the feasibility of
  consistent, available, partition-tolerant web services. SIGACT News
  33(2):51--59

\bibitem[{Gotsman et~al(2016)Gotsman, Yang, Ferreira, Najafzadeh, and
  Shapiro}]{Gotsman:2016}
Gotsman A, Yang H, Ferreira C, Najafzadeh M, Shapiro M (2016) 'cause i'm strong
  enough: Reasoning about consistency choices in distributed systems. In:
  Proceedings of the 43rd Annual ACM SIGPLAN-SIGACT Symposium on Principles of
  Programming Languages, ACM, New York, NY, USA, POPL '16, pp 371--384

\bibitem[{Gray and Reuter(1992)}]{Gray:1992}
Gray J, Reuter A (1992) Transaction Processing: Concepts and Techniques, 1st
  edn. Morgan Kaufmann Publishers Inc., San Francisco, CA, USA

\bibitem[{Gray and Reuter(1993)}]{db:1480}
Gray J, Reuter A (1993) Transaction Processing: Concepts and Techniques. Morgan
  Kaufmann, San Francisco CA, USA

\bibitem[{Guerraoui and Kapalka(2008)}]{Guerraoui:2008}
Guerraoui R, Kapalka M (2008) On the correctness of transactional memory. In:
  Proceedings of the 13th ACM SIGPLAN Symposium on Principles and Practice of
  Parallel Programming, ACM, New York, NY, USA, PPoPP '08, pp 175--184

\bibitem[{Herlihy and Wing(1990)}]{loo:syn:1468}
Herlihy M, Wing J (1990) Linearizability: a correcteness condition for
  concurrent objects. ACM Trans Program Lang Syst 12(3):463--492

\bibitem[{Israeli and Rappoport(1994)}]{Israeli:1994}
Israeli A, Rappoport L (1994) Disjoint-access-parallel implementations of
  strong shared memory primitives. In: Proceedings of the Thirteenth Annual ACM
  Symposium on Principles of Distributed Computing, ACM, New York, NY, USA,
  PODC '94, pp 151--160

\bibitem[{Kemme and Alonso(1998)}]{Kemme:1998}
Kemme B, Alonso G (1998) {A Suite of Database Replication Protocols based on
  Group Communication Primitives}. In: International Conference on Distributed
  Computing Systems, IEEE, IEEE Comput. Soc, pp 156--163

\bibitem[{Ladin et~al(1990)Ladin, Liskov, and Shrira}]{rep:alg:717bis}
Ladin R, Liskov B, Shrira L (1990) Lazy replication: Exploiting the semantics
  of distributed services. In: IEEE Computer Society Technical Committee on
  Operating Systems and Application Environments, {IEEE}, {IEEE} Computer
  Society, vol~4, pp 4--7

\bibitem[{Lamport(1979)}]{Lamport:1979}
Lamport L (1979) How to make a multiprocessor computer that correctly executes
  multiprocess programs. IEEE Trans Comput 28(9):690--691

\bibitem[{Lamport(1986)}]{Lamport:1986}
Lamport L (1986) On interprocess communication. part i: Basic formalism.
  Distributed Computing 1(2):77--85

\bibitem[{Lamport(2006)}]{Lamport:2006}
Lamport L (2006) Lower bounds for asynchronous consensus. Distributed Computing
  19(2):104--125

\bibitem[{Murty(2008)}]{Murty:2008}
Murty J (2008) Programming Amazon Web Services, 1st edn. O'Reilly

\bibitem[{Ozsu and Valduriez(1991)}]{Ozsu:1991}
Ozsu MT, Valduriez P (1991) Principles of Distributed Database Systems.
  Prentice-Hall, Inc., Upper Saddle River, NJ, USA

\bibitem[{Papadimitriou(1979)}]{Papadimitriou:1979}
Papadimitriou CH (1979) The serializability of concurrent database updates. J
  ACM 26(4):631--653

\bibitem[{Peluso et~al(2015)Peluso, Palmieri, Romano, Ravindran, and
  Quaglia}]{Peluso:2015}
Peluso S, Palmieri R, Romano P, Ravindran B, Quaglia F (2015) Disjoint-access
  parallelism: Impossibility, possibility, and cost of transactional memory
  implementations. In: Proceedings of the 2015 ACM Symposium on Principles of
  Distributed Computing, ACM, New York, NY, USA, PODC '15, pp 217--226

\bibitem[{Saeida~Ardekani et~al(2013{\natexlab{a}})Saeida~Ardekani, Sutra, and
  Shapiro}]{rep:syn:sh156}
Saeida~Ardekani M, Sutra P, Shapiro M (2013{\natexlab{a}}) {N}on-{M}onotonic
  {S}napshot {I}solation: scalable and strong consistency for geo-replicated
  transactional systems. In: Symp.\ on Reliable Dist.\ Sys.\ (SRDS), IEEE
  Comp.\ Society, Braga, Portugal, pp 163--172

\bibitem[{Saeida~Ardekani et~al(2013{\natexlab{b}})Saeida~Ardekani, Sutra, and
  Shapiro}]{psutra-europar13}
Saeida~Ardekani M, Sutra P, Shapiro M (2013{\natexlab{b}}) {On the Scalability
  of Snapshot Isolation}. In: Proceedings of the 19th {I}nternational
  {E}uro-{P}ar {C}onference (EUROPAR)

\bibitem[{Saeida~Ardekani et~al(2014)Saeida~Ardekani, Sutra, and
  Shapiro}]{SaeidaArdekani:2014}
Saeida~Ardekani M, Sutra P, Shapiro M (2014) {G-DUR: A Middleware for
  Assembling, Analyzing, and Improving Transactional Protocols}. In:
  Proceedings of the 15th International Middleware Conference, ACM, New York,
  NY, USA, Middleware '14, pp 13--24

\bibitem[{Schiper et~al(2010)Schiper, Sutra, and Pedone}]{psutra-srds10}
Schiper N, Sutra P, Pedone F (2010) P-store: Genuine partial replication in
  wide area networks. In: Proceedings of the 29th IEEE International Symposium
  on Reliable Distributed Systems (SRDS)

\bibitem[{Shapiro et~al(2011)Shapiro, Pregui{\c c}a, Baquero, and
  Zawirski}]{syn:rep:sh143}
Shapiro M, Pregui{\c c}a N, Baquero C, Zawirski M (2011) Conflict-free
  replicated data types. In: D{\'e}fago X, Petit F, Villain V (eds) Int.\
  Symp.\ on Stabilization, Safety, and Security of Dist.\ Sys.\ (SSS),
  {S}pringer-{V}erlag, Grenoble, France, Lecture Notes in Comp.\ Sc., vol 6976,
  pp 386--400

\bibitem[{Shapiro et~al(2016)Shapiro, Saeida~Ardekani, and
  Petri}]{syn:formel:sh190}
Shapiro M, Saeida~Ardekani M, Petri G (2016) Consistency in {3D}. In:
  Desharnais J, Jagadeesan R (eds) Int.\ Conf.\ on Concurrency Theory (CONCUR),
  Schloss Dagstuhl -- Leibniz-Zentrum f{\"u}r Informatik, Dagstuhl Publishing,
  Germany, Qu{\'e}bec, Qu{\'e}bec, Canada, Leibniz Int.\ Proc.\ in Informatics
  (LIPICS), vol~59, pp 3:1--3:14

\bibitem[{Sovran et~al(2011)Sovran, Power, Aguilera, and Li}]{rep:syn:1661}
Sovran Y, Power R, Aguilera MK, Li J (2011) Transactional storage for
  geo-replicated systems. In: Symp.\ on Op.\ Sys.\ Principles (SOSP), Assoc.\
  for Computing Machinery, Cascais, Portugal, pp 385--400

\bibitem[{Terry et~al(1994)Terry, Demers, Petersen, Spreitzer, Theimer, and
  Welch}]{rep:syn:1481}
Terry DB, Demers AJ, Petersen K, Spreitzer MJ, Theimer MM, Welch BB (1994)
  Session guarantees for weakly consistent replicated data. In: Int.\ Conf.\ on
  Para.\ and Dist.\ Info.\ Sys.\ (PDIS), Austin, Texas, USA, pp 140--149

\bibitem[{Viotti and Vukoli\'{c}(2016)}]{Viotti:2016}
Viotti P, Vukoli\'{c} M (2016) Consistency in non-transactional distributed
  storage systems. ACM Comput Surv 49(1):19:1--19:34

\bibitem[{Vogels(2008)}]{rep:syn:pan:1624}
Vogels W (2008) Eventually consistent. {ACM} {Q}ueue 6(6):14--19

\bibitem[{Wang et~al(2014)Wang, Talmage, Lee, and Welch}]{Welch:2014}
Wang J, Talmage E, Lee H, Welch JL (2014) Improved time bounds for linearizable
  implementations of abstract data types. In: 2014 IEEE 28th International
  Parallel and Distributed Processing Symposium, pp 691--701

\bibitem[{Wiesmann and Schiper(2005)}]{Wiesmann:2005}
Wiesmann M, Schiper A (2005) {Comparison of database replication techniques
  based on total order broadcast}. IEEE Transactions on Knowledge and Data
  Engineering 17(4):551--566

\bibitem[{Zemke(2012)}]{Zemke:2012}
Zemke F (2012) What.s new in sql:2011. SIGMOD Rec 41(1):67--73

\end{thebibliography}

\section{Cross-References}

\begin{compactitem}
\item
  Achieving LowLatency Transactions for Geo-Replicated Storage with Blotter
\item
  Conflict-free Replicated Data Types (CRDTs)
\item
  Coordination Avoidance
\item
  Data replication and encoding
\item
  Databases as a service
\item
  Geo-replication Models
\item
  Geo-scale Transaction Processing
\item
  In-memory Transactions
\item
  NoSQL Database Principles
\item
  Storage hierarchies for big data
\item
  TARDiS: A Branch-and-Merge Approach To Weak Consistency
\item
  Transactional Middleware
\item
  Weaker Consistency Models{\slash}Eventual Consistency.
\end{compactitem}

\end{document}